# Role Of Polymer Loops In DNA Replication

Suckjoon Jun and John Bechhoefer

Department of Physics, Simon Fraser University, Burnaby, British Columbia, V5A 1S6

tel. 604-291-5924; fax: 604-291-3592; e-mail: johnb@sfu.ca;
http://www.sfu.ca/chaos/

**ABSTRACT**

Loop formation in long molecules occurs many places in nature, from solutions of carbon nanotubes to polymers inside a cell. In this article, we review theoretical studies of the static and dynamic properties of polymer loops. For example, long polymers must search many configurations to find a "target" binding site, while short polymers are stiff and resist bending. In between, there is an optimal loop size, which balances the entropy of long loops against the energetic cost of short loops. We show that such simple pictures of loop formation can explain several long-standing observations in DNA replication, quantitatively.

## INTRODUCTION

Over the last decade, many new, exciting experimental techniques to manipulate single molecules have been developed that have renewed interest in the classical theory of single-chain polymers [1-3]. One case, the looping of polymers, has important applications of current interest. For example, Sano *et al.* [4] used covalent reactions to produce ring-shaped carbon nanotubes. In biology, the long molecule *par excellence* is DNA, and looping in DNA is important in several different contexts. For example, Goddard *et al.* [5] studied the opening and closing dynamics of short, single-stranded DNA fragments ("molecular beacons"), finding that their rigidity depends strongly on the specific nucleotide sequence. The sensitivity to sequence can be used to diagnose rapidly diseases that correlate with changes in a single, specific nucleotide in a genome ("single-nucleotide polymorphism," or SNP) [6]. In living cells, at scales of hundreds of bases, DNA looping plays an important role in gene regulation, where, for example, it can multiply greatly enzyme reaction rates by allowing a DNA-bound protein to interact with a target site on the same DNA molecule much more frequently than a free protein would [7, 8]. Loops may also appear in complex DNA-protein structures, such as the 30-nm chromatin fibre, at scales of thousands of bases, or even longer [9]. Finally, in a non-DNA example, the looping of an amino acid chain is one of the key issues in protein folding; the resulting "loop regions" often form the binding sites of other molecules [10].

All of these examples involve long molecules with an intrinsic stiffness. In this article, we review some of the theoretical studies done on polymer looping, paying particular attention to the role of stiffness in defining a characteristic loop size. We then show that models incorporating the effects of looping can help in interpreting experimental data on DNA replication.

## THEORETICAL APPROACHES TO MODELING POLYMERS

### Configurations

We first review the overall classification of polymer models, both discrete and continuous. The simplest discrete polymer model is the freely jointed chain (FJC). Fig. 1A shows a model FJC as a chain of freely joined vectors of fixed length $b_0$. The FJC ignores both monomer interactions and finite chain stiffness and can be thought of as a random walk of a fixed step length, where each step is independent of the previous trajectory. Usually, the "size" of a polymer chains is defined as $\sqrt{\langle \vec{R}^2 \rangle}$, and one can derive a very simple scaling law $\sqrt{\langle \vec{R}^2 \rangle} \propto N^{\frac{1}{2}}$ from

**(1)**

$$\langle \vec{R}^2 \rangle = \left\langle \sum_i \Delta \vec{r}_i \ \sum_j \Delta \vec{r}_j \right\rangle = \sum_i \langle \Delta \vec{r}_i^2 \rangle + \sum_{i \neq j} \langle \Delta \vec{r}_i \cdot \Delta \vec{r}_j \rangle = N b_0^2$$

When $N \to \infty$, the distribution of end-to-end vectors $\vec{R}$ is Gaussian. In a variant of the FJC, beads are separated by freely jointed linear springs, which leads to a Gaussian distribution of bond lengths. For large $N$, the FJC and this "Gaussian-chain" model behave identically.

A more realistic discrete model of polymer, the freely rotating chain (FRC), is shown in Fig. 1B. The FRC consists of vectors with fixed bond angle, but with completely free dihedral angles, thus naturally incorporating a finite stiffness. In the FRC, $\langle \vec{R}^2 \rangle$ can be calculated exactly in a straightforward way, and it is easy to show that the FRC also follows the same scaling law in $N$ as the FJC. Next, we define a quantity called the "persistence length" as

**(2)**

$$l_p = \lim_{N \to \infty} \left\langle \vec{R} \propto \vec{u}_0 \right\rangle = \frac{b_0}{1 - \cos\theta},$$

which is the average length of the projection of the end-to-end vector $\vec{R}$ along the direction of the first bond vector. As we shall show below, the persistence length is a measure of chain stiffness.

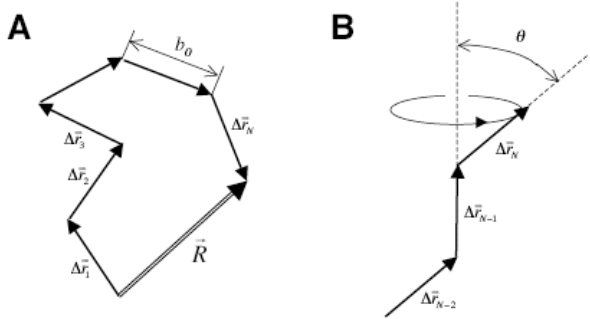

**Figure 1.** Discrete models of polymer. (A) Freely Jointed Chain (FJC) (B) Freely rotating chain.

The continuous limit of the FRC is the Kraty-Porod (KP) wormlike chain [11]. We define the total contour length $L = N \cdot b_0$ and the contour distance $s$ ($0 \le s \le L$) from the zero'th to the $i$'th vector by $s = i \cdot b_0$. We then take the limit $N \to \infty$, $b_0 \to 0$, and $\theta \to 0$, with constraints that the chain length $L$ and the persistence length $l_p$ remain constant. The discrete chain contour then becomes a continuous, differentiable space curve. The statistical properties of the KP wormlike chain are determined by an effective free energy

**(3)**

$$H = \frac{\kappa}{2} \int_0^L \left( \frac{\partial \vec{u}(s)}{\partial s} \right)^2 ds, \quad \text{with} \quad |\vec{u}(s)| = 1,$$

where $\kappa \equiv l_p * k_B T$ is the bending modulus of the polymer, and the unit tangent vector $\vec{u}(s)$ at $s$ on the curve is defined as $\vec{u}(s) = \frac{d\vec{r}(s)}{ds}$, with $\vec{r}(s)$ is the position vector. We note that imposing the constraint of fixed polymer length, $|\vec{u}(s)| = 1$, is one of the major difficulties in handling the model analytically [12].

Several quantities, nonetheless, are known exactly. One of the most important is the spatial correlation function for unit tangent vectors [13],

**(4)**

$$\left\langle \vec{u}(s) \quad \vec{u}(s') \right\rangle = \exp\left(-\frac{|s - s'|}{l_p}\right).$$

Using Eq. 4, we can also calculate $\left\langle \vec{R}^2 \right\rangle$ exactly,

**(5)**

$$\left\langle \vec{R}^2 \right\rangle = \int_0^L \int_0^L \left\langle \vec{u}(s) \quad \vec{u}(s') \right\rangle ds ds' = 2l_p L - 2l_p^2 \cdot \left(1 - e^{-\frac{L}{l_p}}\right).$$

In Eq. 5. for $L \ll l_p$, we have $\left\langle \vec{R}^2 \right\rangle = L^2$: the rod is rigid. For $L \gg l_p$, we have $\left\langle \vec{R}^2 \right\rangle = 2l_p L$, which is identical to Eq. 1 if we identify $b_0' = 2l_p$ and $N' = L/b_0'$ as effective segment lengths and polymerization indices, respectively. The behaviour in these two limits shows that the KP wormlike chain interpolates between the rigid rod and the Gaussian chain. Hence, the persistence length $l_p$ is a measure of the chain stiffness in the KP model. One often uses a dimensionless chain length, $l = L/l_p$.

We note that neither the KP nor the lattice models considers the torsional energy of a chain, which can lead to complications such as supercoiling and knotting [14]. The helical wormlike (HW) chain model has both bending and torsional energies, and it has been very successful in applications involving short lengths of DNA. Formally, the HW chain is obtained from a discrete chain with coupled rotations (the dihedral-angle distributions are non-uniform) [15].

**The Looping Problem**

We now turn our attention to the looping problem for semiflexible polymers. Imagine a dilute solution of semiflexible polymers with two sticky ends in equilibrium (Fig. 2). (For dilute solutions, one can ignore inter-chain reactions.) For simplicity, we also assume the distribution of contour lengths of polymers to be uniform. In single-polymer-chain problems, a central quantity for characterizing chain conformations is the distribution function of the end-to-end vector $\vec{R}$, $G(\vec{R};L)$, where $L$ is the total contour length. Then $G(\vec{R};L)$ can be identified with the probability density for finding the two ends of the polymer chain separated by $\vec{R}$ and $G(\vec{R}=0;L)$ with the ring-closure probability [16]. If we consider $G(R=0; L)$ as a function of $L$, it is the number of closed-chain configurations for a polymer of length $L$.

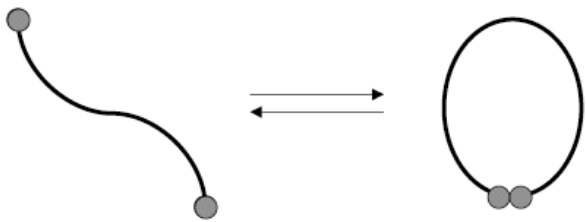

**Figure 2.** A semiflexible polymer with two sticky ends in a dilute solution. Here, we assume that the end-to-end interaction is short ranged, so that the two ends react only if the two ends meet by diffusion.

Because of the intrinsic stiffness of the KP chain, $G(\vec{R}=0;L)$ shows a completely different behaviour from that of Gaussian chains, where entropy determines all. Loops of semiflexible chains cannot have arbitrary sizes: It costs too much energy to bend a stiff polymer into a loop whose length is comparable to the persistence length $l_p$, thus favoring larger loops. On the other hand, as loop sizes increase, a larger configuration space must be searched, thus favoring shorter loops. The optimal loop size, which balances energy and entropy, is 3-4 times the persistence length.

For the ideal phantom FJC with no stiffness, the end-to-end distribution function $G(\vec{R};L)$ is known exactly and converges to a Gaussian distribution in the large-$N$ limit, with mean-square average end-to-end distance $\langle \vec{R}^2 \rangle = N b_0^2$ or $2Ll_p$ (see earlier definition below Eq. 5). The "persistence length" $l_p$ is independent of temperature because its microscopic origin lies in steric constraints rather than in the bending rigidity of the backbone [15]. The effect of excluded volume or self-avoidance profoundly changes the properties of flexible chains, and its analytical treatment is difficult. For a semiflexible polymer whose size is comparable to the persistence length, however, we can reasonably ignore the effect of excluded volume, which become quantitatively important only for long chains. Given the long history of the KP wormlike chain, it is surprising that a good approximation to $G(\vec{R};L)$ for $L \cong l_p$ was obtained only recently by Wilhelm and Frey [17], and then in a slightly different form by Thirumalai and Ha [18] (Note: for $L \geq 10l_p$, $G(\vec{R};L)$ is well-described by the Daniels approximation, which is an asymptotic expansion about a Gaussian distribution [19]). Previous authors had started from chains near the rod limit, expanding in powers of $l = L / 2l_p$ to obtain corrections. Although these calculations are quite involved, they do not produce even qualitatively correct results. For the ring-closure probability $G(R = 0; L)$, however, we note that Yamakawa and Stockmayer have obtained an approximate expression that is qualitatively correct, using the KP model [20]. Later, Shimada and Yamakawa extended their results to the helical wormlike chain model [21], deriving an often-used approximate expression for the ring-closure probability, valid for $L < 10\, l_p$ [22],

**(6)**

$$G\left(R=0;l\right) = 28.01 \cdot l^{-5} \cdot \exp\left(-\frac{7.027}{l} + 0.492 \cdot l\right)$$

In Eq. 6, the different configurations of dihedral angles have been averaged over.

### Dynamics Of Loop Formation

In the previous section, we described the conformations of polymers in thermodynamic equilibrium. In biology, the only systems in equilibrium are dead, and one is forced to think also about non-equilibrium conditions and about dynamics. For loops, we have focused on $G\left(\vec{R}=0;L\right)$, the number of loop configurations for a polymer of length $L$. Often, what we really want to know is the rate of loop formation and other similar questions. We can imagine two limiting cases. Let us first imagine that when the two ends of the polymer meet, the probability of sticking (binding) is very low. The polymer has ample time to sample all available configurations before actually binding. (We assume that the binding, once it occurs, is irreversible. The opposite case, where the loop opens and closes many times, is termed a "living polymer" [23].) The time to form a loop would then be approximately

**(7)**

$$\tau_R\left(L\right) = \frac{\int dR\, G\left(R;L\right)}{G\left(R=0;L\right)} \cdot \eta_0^{-1} \cdot \tau_0$$

where the $G$-factors are the number of non-loop conformations per loop conformation, $\eta_0$ is the probability for ends to stick per encounter, and $\tau_0$ is the time to explore one configuration. In this "quasistatic" picture, the $G$-factors are obtained from static, equilibrium theory, and dynamics enters only in the calculation of $\tau_0$. In analogy with other chemical processes, this is often known as the reaction-controlled limit.

Alternatively, if the sticking probability $\eta_0 \approx 1$, a loop will form the very first time the ends meet. The time to form a loop, $\tau_D$, is then a "first-passage time:" it is the time randomly diffusing ends require to encounter each other. This is often termed the diffusion-controlled limit (see, for example, [24] and references therein).

Most work on biological systems assumes, often implicitly and usually without much justification, that the first limit pertains. In our own work, below, we follow that tradition, but it is useful to briefly sketch some of the theoretical approaches to "true" dynamics.

The overall framework begins with a model for the dynamics of the polymer chain. The traditional starting point is the Rouse model, which is a Gaussian chain with the beads subject to local friction, in the strongly overdamped limit. In the Zimm model, the beads interact with their nearest neighbours via springs and with more distant neighbours hydrodynamically: the motion of one bead stirs the fluid, which exerts a force on distant beads. The Rouse and Zimm models can be used to estimate $\tau_0$ in Eq. 7. (For dilute solutions, the Zimm model does a much better job.)

In the diffusion-controlled limit, Wilemski and Fixman (WF) have extended the Rouse model [25]. In the latter, the polymer's state "diffuses" through configuration space. In their extension, WF add a "sink" that freezes the polymer's state when it wanders into a loop configuration. Unfortunately, Rouse models are not quantitatively accurate for dilute polymer solutions, and the better Zimm calculation has not been done.

A cruder way of looking at the dynamics of loop formation is to group all of the open configurations together, giving a "two-state" system, where the polymer is either in a "linear" or "loop" state (for flexible polymers, see [26] and references therein). At time $t = 0$, we assume that the conformations of all the polymers in the solution are linear. If the end-sticking reaction is irreversible, all the polymers will eventually form loops even if there is a bending energy (Fig. 3A). In other words, there is a transition from one state (linear) to another (loop) due to thermal fluctuations, which enables the stiff polymer to overcome the bending energy. This problem is equivalent to the well-known Kramers problem in non-equilibrium statistical mechanics [27], where a particle is trapped in a potential well but thermal fluctuations provide an energy source that allows the trapped particle to escape (Fig. 3B). If one considers the rigid-rod limit of the KP chain, the effective potential energy becomes $E_{eff} = \dfrac{\kappa}{2} \dfrac{\theta^2}{L}$, where $\kappa$ is the bending modulus of the stiff chain and $L$ is the total contour length (Fig. 3C). One can then show that the "escape" rate $r = \tau_D^{-1}$ is

**(8)**
$$r = J(\theta \approx 2\pi) \approx f(l, \mathbf{C}) \cdot e^{-\frac{C'}{l}},$$

where $J$ is the current of right-moving probability and $C'$ is a constant factor. The prefactor $f$ is a slowly varying function of a reduced chain length $l$ and a set of constants $\mathbf{C}$ that includes the friction coefficient $\gamma$. Eq. 8 predicts a rapid suppression of escape rate for small $l$. Since $\bar{\tau}$ increases monotonically as $l^2$ [28] for large $l$, (flexible-chain limit), by interpolating these two extreme cases, one still expects an optimum value of $l > 0$ for the maximum escape rate or minimum loop-closure time. Thus, even in a dynamical picture, one expects qualitatively the same picture as in the quasistatic case: loops are suppressed for $L \approx l_p$, form most rapidly for $L \approx$ few $l_p$, with the probability tailing off for longer $L$'s. This equivalence justifies our use of the quasistatic limit, where, if we are merely interested in relative numbers of loops of different sizes, the factor $\tau_0$ is roughly constant and only static information is required.

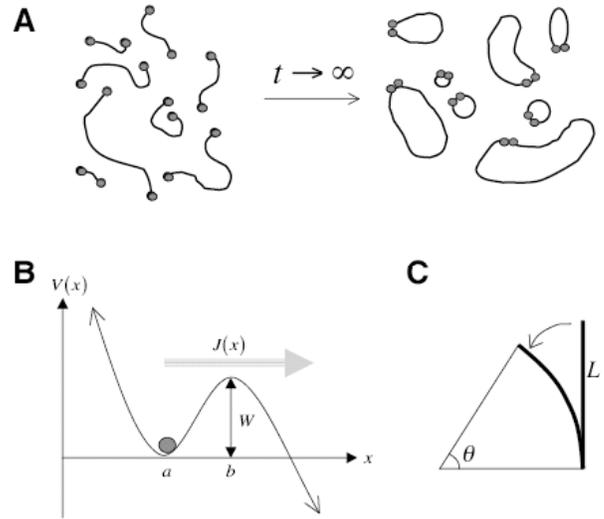

**Figure 3.** Two-state model of loop formation and the Kramers escape problem. (A) Semiflexible polymers in a dilute solution. The reaction is irreversible and eventually all the linear polymers in the solution form loops. (B) A classical (not quantum) Brownian particle is trapped in a potential well of height $W$. Because of the thermal energy, the particle has a non-zero probability to escape from the well. $J(x)$ is the flux of right moving probability. (C) Bending of a rigid rod. A rigid rod has to overcome a "potential" energy $E_{eff} = \dfrac{\kappa}{2} \dfrac{\theta^2}{L}$ in order to bend, where $\kappa$ is the bending modulus.

## EXPERIMENTAL APPLICATION TO DNA REPLICATION

The recent complete sequencing of the human genome has stimulated great interest in modeling various aspects of DNA. The function of DNA is to store genetic information in a way that can be interpreted by the cell and replicated for future generations. The amount of information is large: the human genome, for example, consists of about $3*10^9$ nucleotide base pairs (bp) (We note that lengths of DNA are often measured in base pairs: 1 Mb (megabases) = $10^3$ kb (kilobases) = $10^6$ b (bases) or bp). Cells have very sophisticated micro-machineries to copy this information quickly and accurately. Although the organization of the genome for DNA replication varies considerably from species to species, the duplication of most eukaryotic genomes shares a number of common features:

(1) DNA replication starts at a large number of sites, "origins of replication." The amount of DNA replicated from each origin is known, informally, as an "eye" because of its appearance in electron microscopy.

(2) The position of each *potential* origin that is "competent" to initiate DNA replication is determined before the beginning of the synthesis part of the cell cycle ("*S*-phase"), when several proteins, including the origin recognition complex (ORC) bind to DNA, forming a pre-replication complex (pre-RC).

(3) During *S*-phase, a particular potential origin may or may not be activated. Each origin is activated not more than once during the cell-division cycle.

(4) DNA synthesis propagates at replication forks bidirectionally with propagation speed or fork velocity *v*, from each activated origin.

(5) DNA synthesis stops when two newly replicated regions of DNA meet.

Processes (3)-(5), depicted in Fig. 4A, can be described by a stochastic model that is formally equivalent to nucleation-and-growth models developed in the 1930s [29-31]. The "Kolmogorov-Johnson-Mehl-Avrami" (KJMA) model has since been widely used by metallurgists and other scientists to analyze the freezing of liquids. To understand the model, consider an ice-cube tray that you fill with water and put in a freezer. Some time later, it is all frozen. In the meantime, three processes have occurred:

(1) nucleation of discrete solid domains (analogous to the replicated regions);
(2) growth of the domains;
(3) coalescence when two expanding domains meet.

Of course, in the case of DNA, the "freezing" is one dimensional. But that makes the model much easier to solve, and, indeed, theoretical physicists in the 1980s and 1990s had devoted some effort to the one-dimensional KJMA model [32, 33], not because there was any real application but because one could pursue analytic and numerical work much further than in three dimensions!

Inspired by this analogy, we applied the KJMA model to data from the recent "molecular combing" [34] experiment by Herrick *et al*. [35-37]. These experiments used two-colour fluorescent labeling of DNA bases. One begins (in a test tube, alas) by labeling the bases used in replicating the DNA with, say, a red dye. At some point, during the replication process, one floods the test tube with green-labeled bases and allows the replication cycle to go to completion. One then chops up the DNA into fragments that are examined under a microscope. Regions that replicated before adding the dye are red, while those labeled afterwards are predominantly green. The alternating red-and-green regions form a kind of snapshot of the replication state of the DNA fragment at the time the second dye was added. Varying that time in different runs allows one to systematically look at the progression of replication throughout *S*-phase. From images of these DNA fragments, we can form statistics about replicated lengths ("eyes"), non-replicated lengths ("holes"), and distances between origins ("eye-to-eye" lengths). Our problem then becomes the reverse of the usual application of the KJMA to problems in materials science. In the latter, one generally knows the temperature of the material, and hence the nucleation rate, as a function of time, and one wishes to predict the fraction solidified, the size of solid domains, etc. In our case, we have information about the eyes, holes, and eye-to-eye lengths, and we wish to infer the rates of origin initiation $I(t)$ throughout $S$-phase.

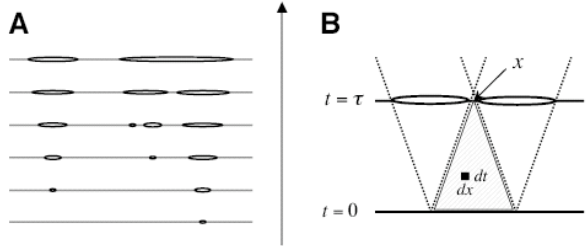

**Figure 4.** Space-time diagram of one-dimensional nucleation and growth. (A) The replicated/solid domain is shown as a bubble, or "eye". (B) The probability for a point $x$ at time $\tau$ to remain unreplicated is determined by the requirement that there be no initiations in the shaded area. Here, $I(t)$ is the initiation probability per unit time per unit unreplicated length. See Eq. 9.

Fortunately, it is possible to "invert" the KJMA model, so that from experimental data such as the average length of replicated fragments as a function of the fraction $f$ of total DNA replicated (Fig. 5A), we could extract various parameters that govern the kinetics of DNA replication in early-embryo *Xenopus laevis*, a type of frog often used in developmental studies [36]. In particular, the frequency of origin firings $I(t)$ and the fork velocity $v$ were determined directly from the data (Fig. 5B). Once we know $I(t)$ and $v$, almost all quantities such as the mean sizes of replicated and unreplicated domains can be calculated either analytically or numerically. For example, from the space-time diagram in Fig. 4B, the fraction of DNA $f$ replicated a time $\tau$ after the start of $S$-phase is given by

**(9)**
$$f(\tau) = 1 - \prod_{x,t \in \Delta} \left(1 - I(t) \, dx \, dt\right)$$
$$= 1 - e^{-v \int_0^\tau I(t)(\tau-t)dt}$$

Notice that the extracted $I(t)$ gives information about the time evolution of the overall origin initiation rate but not about the origin distribution along the genome. The lack of spatial information reflects the nature of the kinetic model used, where all spatial information was averaged out to create a "mean-field," spatially homogeneous model. But are the initiations of individual origins independent of each other? Does origin activation enhance and/or suppress initiation of its neighboring origins?

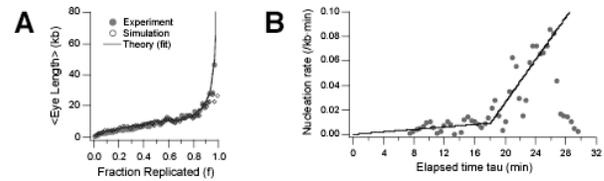

**Figure 5.** Inverting the KJMA model. (A) <eye size> vs. fraction replicated $f$. (B) One can apply the KJMA model to extract (invert) the spatially averaged ("mean-field") probability of initiation $I(t)$ from the data shown in (A).

To answer such questions, we need to understand more about the biology of replication. In eukaryotic organisms, origin initiation occurs in several steps:

1. In the $G1$-phase of the cell cycle, several proteins (ORC complexes, Cdc6, Cdt1) bind to the DNA.
2. Each of the attached ORC-Cdc6-Cdt1 complexes recruits 10-40 copies of the six proteins MCM2-7. They then also recruit the proteins Cdc6 and Cdt1, forming a pre-replication complex (pre-RC).
3. In $S$-phase, the pre-RC is activated via a complex sequence of protein interactions, allowing MCM2-7 to unwind the DNA to start replication;
4. DNA replication is initiated (at most once each cycle) at (some of) the pre-RCs.

Thus, we can sharpen our questions: Since initiation must occur at one of the pre-

RC sites, we may ask, How are pre-RCs distributed along the genome? Do the pre-RCs initiate independently, or is there a correlation between the activities of neighboring sites?

In *Xenopus* early embryos, the assembly of pre-RC is DNA sequence independent [38], but the amount of pre-RC and the nature of its distribution along the genome is controversial [39, 40]. Because the duration of *S*-phase ($\approx$ 20 min.) is determined by the largest separation between pre-RCs and the replication-fork velocity ($\approx$ 10 bp/sec), this separation has an absolute upper limit ($\approx$ 20 kb for the *Xenopus*); otherwise, one has the "random-completion" problem [41]: Even a single large gap between origins will lengthen *S*-phase, leading to a usually fatal abnormality in the developing organism. The limit is all the more serious in that many of the mechanisms that normally allow a cell to pause to allow unreplicated areas to "catch up" are not present in the early embryo.

Currently, there are two different views on how to resolve the random-completion problem. The first is that the density of pre-RCs is roughly the same as that of activated origins [40]. "This would require some regularity in pre-RC spacing, as well as almost 100% initiation efficiency.

The second is that the distribution of pre-RC sites is random and that there are many more pre-RC sites than origins [39, 42]. In this case, the separation between pre-RC sites should follow an exponential distribution [43]. The average density of sites must be high enough that the long tail of large separations has only a very small chance of producing a fatally large origin gap.

In both scenarios presented above, many people have speculated that the higher-order structure of the DNA-protein complexes that make up chromosomes ("chromatin") can have loops and that these loops can impose some regularity to the separation of replication origins [39, 40, 44]. Here, we list some observations that look somewhat scattered and unrelated at first glance, but can all be explained and related as consequences of chromatin looping, providing a natural solution to the random-completion problem:

(1) Initiations are inhibited for distances $\leq$ 2-4 kb [39].

(2) Most initiations are separated by 5-15 kb on *Xenopus* sperm chromatin [40]. In a similar system, early-embryo *Drosophila* (a fruit fly), the average origin spacing is about 7.9 kb [45].

(3) There is a weak but statistically significant positive correlation between the sizes of neighboring replicated domains. Large eyes tend to have large neighbours [40].

(4) Although one's first guess might be that the proteins involved in DNA replication would find the DNA and move along it, there is evidence that the reverse is actually the case: Proteins such as the polymerases required for replication are localized in "replication factories" at discrete sites in the cell nucleus, and the DNA comes to them [46].

(5) In eukaryotic organisms, chromosomes have several levels of higher-order structure, each with its own length scale. The various structures can be modeled as wormlike chains. In particular, the persistence length of the "30-nm" fiber is about 30 nm, which corresponds to 2-4 kb of DNA [8, 47, 48].

As we shall argue, these points are natural consequences of the looping scenario. In order to show this, we performed Monte-Carlo simulations of the replication process. One can think of DNA as one-dimensional lattice (just like the 1-D Ising model with two possible spin values, i.e., up and down), where each lattice site is assigned either 1 or 0, depending on whether the site has been replicated or not, respectively. For example, at time $t$=0, all the lattice sites are 0. One can then convert $I(t)$ defined earlier to the mean probability of initiation $\overline{p}(t) = I(t) \cdot \Delta x \cdot \Delta t$ at site $x$ and time $t$: if $\overline{p}(t)$ is larger than a random positive real number smaller than 1 (standard Monte-Carlo step), there is an initiation and the lattice value changes from 0 to 1. Once an initiation occurs, the replicated domain (1's) grows bidirectionally with fork velocity $v$. In the simulations, each lattice is updated with a typical timestep $\Delta t = 0.2$ min. The natural size of lattice then would be $v \cdot \Delta t$, which is about 120 bp for the measured fork velocity $v = 600$ bp/min. The lattice scale is then roughly the size of pre-RC. We sample the simulation results at the same time points as the actual experiments ($t = 25, 29, 32, 35, 39,$ 45 min.) Each sampled molecule is cut at random site to simulate the combing process. The lattice is then "coarse grained" by averaging over several pixels (typically four) in order to take into account the resolution of scanned optical images, which is about 0.24 $\mu$m = 0.48 kb. Finally, the coarse-grained fragments were analyzed to compile statistics concerning replicon sizes, eye-to-eye sizes, etc. that were directly compared to experimental data [36]. One can refine the simulation to include a site-specific probability of initiation $p(x, t)$.

It is easy to show that any scheme where the probability of initiation depends only on time and location along the genome cannot give the kind of correlations noted in point (3), above. The basic idea is that the independence of initiation events carries through to the sizes of domains. Only when initiations "know" about prior events (i.e., prior initiations) do correlations emerge.

This led us to consider other scenarios, the most natural of which is one where a replication factory "captures" the chromatin (structured DNA) at some point along its length. Neighboring origin sites, which are defined by ORCs assembled onto the genome, can then loop back to the same factory for initiation. We assume that the loop-formation dynamics are quasistatic, so that their size distribution is governed by Eq. 6. The looping scenario is easy to incorporate into the Monte-Carlo simulations.

As we have seen, loops of chromatin cannot have an arbitrary size but should be 3-4 times $l_p$ (Eq. 6 implies a peak probability at $L = 3.4*l_p$). This translates into an expected enhancement of initiations at origin separations of 6 to 16 kb and a suppression for smaller separations -- just what is observed!

Moreover, when we used the loop-formation scenario in the simulation as a small perturbation to the mean-field initiation frequency $I(t)$, we obtained a positive correlation consistent with previously reported value C = 0.16 by Blow *et al*. We could also reproduce the distribution of separations between neighboring replicated domains. While we do not have space to show the detailed comparison of the models [49], we give one example. Fig. 6 shows histograms that record the relative position inside a domain of unreplicated DNA, or hole, of new origins. Fig. 6A shows the distribution for small holes, 8-22 kb in length [50]. The experimental data shows a strong peak near 0.5, implying a tendency for origins to be as far away from other replicating domains as possible. By contrast, the experimental data for large holes shows a much more uniform distribution. In simulations that use random initiation, new origins can appear almost anywhere in a hole, regardless of its size. This picture fits the large-hole data (Fig. 6B) but not the small-hole data (6A). By contrast, when we put in the effects of looping, which lead to a suppression of origin initiation at very close spacings and an enhancement of initiation at a larger, characteristic distance, the simulation results match more closely the data of Fig. 6A, while continuing to agree with the large-hole case [49].

Thus, by considering the effects of chromatin loops in the context of our model of DNA replication, we can explain the typical size of replicated domains, the inhibition of origin initiation near already activated origins, and the correlation in the initiation times of neighbouring active origins. The regularity of origin spacings imposed, along with the increasing $I(t)$ in Fig. 5B, solve, in a natural way, the random-completion problem. Finally, in a more stringent test, we can also fit the detailed distributions from the *Xenopus* experiment. Models with purely random origin activation give results that do not agree with those distributions [49].

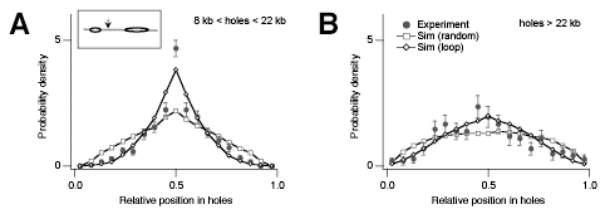

**Figure 6.** (A) Histogram of positions of initiation events for holes 8-22 kb in size. The events are determined by looking for replicated domains that are small enough to be very likely to contain only a single replication origin. The state of the molecular fragment is then propagated back in time to the moment of initiation, where one records the hole size and relative position of the initiation event within the hole. The inset shows a hole flanked by two eyes. The experimental histogram shows that it is more likely that a new initiation occurs near the centre of a hole, an observation compatible with the looping scenario but not with the purely random initiation scenario. (B) Holes larger than 22 kb. The difference between experiment and simulations (both random and loop formation) is much smaller than for small holes in (A).

## CONCLUSION

We have described above how the statistical mechanics of polymer looping can help explain a number of observations about DNA replication. As always, when two fields touch, their meeting is not the end, but the beginning of a long story. Here, we can see two kinds of further development. The first extends from polymer physics to biology, where many other single-molecule experiments are giving detailed information and where the ideas described above may be applied. The molecular beacons described in the introduction would be one such example. The detailed application of polymer theories also suggests specific ways to conduct the experiments so that the right information is captured.

It is worth dwelling on the importance of single-molecule experiments. Traditional work in biology, biochemistry, and biophysics uses solutions of molecules. Such experiments give information about the *average* behaviour of the studied molecules. The great virtue of single-molecule experiments is that they collect information about *individual* molecules. Because noise is always important at molecular scales, the behaviour of one individual molecule will differ from the next. One must then study many such molecules, forming probability densities for the quantities of interest. In the *Xenopus* experiments, for example, information was collected not just on the *average* eye length but on the *distribution* of eyes, as well. This kind of detailed result allowed us to distinguish between the random-initiation and looping models. Both models can fit data such as the curve of average eye size (Fig. 4A), but their distributions differ. We note, too, that noise, probability distributions, and correlations are the natural business of statistical mechanics, making it the right way to attack such problems.

The second extension of our work is that, as is so often the case, the concrete situations found in biological systems lead naturally to new physical problems. For example, all studies of biological looping problems to date have implicitly assumed the "quasistatic" dynamics approximation described above. There has been little careful thought as to when such a limit applies and when the dynamical, "first-passage-time" limit is more appropriate. As we have seen, few results are available for "true" dynamics.

Even in the quasistatic case, there are further problems to pursue. For example, in the replication-factory picture, the DNA is initially trapped at one site. If a long piece loops back, the large loop may subdivide into smaller subloops. Such "multiple looping" has not yet been considered theoretically. There is much work to be done!

## ACKNOWLEDGEMENTS

We thank Aaron Bensimon and John Herrick for their generous collaboration and Philippe Pasero for helpful comments. This work was funded by NSERC.